\documentclass[11pt, a4paper]{article}
\pdfoutput=1
\usepackage{jheppub}
\usepackage{graphicx}
\usepackage{epsfig,amsmath,amsthm}
\usepackage{epstopdf}

\newcommand{\be}{\begin{equation}}
\newcommand{\ee}{\end{equation}}
\newcommand{\bea}{\begin{eqnarray}}
\newcommand{\eea}{\end{eqnarray}}
\def\bse{\begin{subequations}}
\def\ese{\end{subequations}}

\def\IZ{\relax\ifmmode\hbox{Z\kern-.4em Z}\else{Z\kern-.4em Z}\fi}

\newcommand{\non}{\nonumber \\}

\def\del{{\partial}}

  \def\eps{\epsilon}

\def\lam{\lambda}

\def\presub{\vspace{.5cm} \noindent}

\def\bi{\begin{itemize}} \def\ei{\end{itemize}}

\def\({\left(} \def\){\right)}
\def\[{\left[} \def\]{\right]}
\def\<{\left<} \def\>{\right>}

\title{Algebraic aspects of when and how a Feynman diagram reduces to simpler ones}

\author{Barak Kol  \\
{\it Racah Institute of Physics, Hebrew University, Jerusalem 91904, Israel} \\
{\tt barak.kol@mail.huji.ac.il}
}

\abstract{The method of Symmetries of Feynman Integrals defines for any Feynman diagram a set of partial differential equations. On some locus in parameter space the equations imply that the diagram can be reduced to a linear combination of simpler diagrams. This paper provides a systematic method to determine this locus and the associated reduction through an algebraic method involving factorization of maximal minors.
}

\begin{document}

\maketitle

\section{Introduction}
\label{sec:intro}

Feynman diagrams and the associated integrals are at the computational core of quantum field theory and their evaluation attracts considerable attention see e..g \cite{SmirnovBooks}. Two important and well-known methods for their computation are Integration By Parts (IPB) \cite{ChetyrkinTkachov1981} and Differential Equations (DE) \cite{DE}. The more recent method called Symmetries of Feynman Integrals (SFI) \cite{SFI} is closely related to them. It considers a Feynman integral $I$ associated with a diagram of fixed topology yet most general parameters, namely masses and kinematical invariants denoted collectively by $X$. Given a diagram SFI defines a set of partial differential equations in $X$, with the idea that the integral can be determined as a solution of this equation set (at least partially), rather than through direct integration.  

The method was applied to the two-loop vacuum diagram (or ``diameter diagram") \cite{SFI, locus}, the 1-loop propagator diagram (``bubble") \cite{bubble} and most recently to the vacuum seagull diagram, a certain 3 loop vacuum diagram, where novel evaluations were obtained for certain 3 mass scales. 

The SFI method has aspects of geometrical symmetry as the equation set defines a continuous group $G$ which acts on parameter space $X$ and had a rather direct geometrical interpretation. More specifically, $G$ consists of all possible linear redefinitions of the loop currents and external momenta which preserve the propagator hyperspace in the space of quadratics in currents, see \cite{locus,VacSeagull} for definitions.

Generically, on orbits of $G$ in $X$ SFI reduces the computation of the Feynman integral to a line integral over simpler diagrams (up to some boundary or initial condition such as a base point).  However, at some locus in $X$ the equations set is singular and produces algebraic rather than differential equations \cite{locus}. On some components of this locus the algebraic equation allows to reduce $I$ into a linear combination of simpler diagrams (instead of a line integral).

The purpose of this note is to present algebraic methods for determining the singular locus and when relevant, the reduction to simpler diagrams. We start in section \ref{sec:prelim} by recalling from linear algebra the role of maximal minors and their decomposition as well as the LU decomposition. In section \ref{sec:reduction} we proceed to apply a generalization of these methods to the SFI equation set and to obtain our results. In section \ref{sec:demon} the algebraic methods are demonstrated on the diameter diagram. We conclude with a discussion in section \ref{sec:disc}.

\section{Linear algebra preliminaries}
\label{sec:prelim}

In this section we review maximal minors and the LU decomposition in linear algebra. This will serve as a basis for the next section.

\subsection*{Warm up}

As a first warm up consider an $n+1$ by $n$ matrix of rank $n$ (the maximal possible rank) \bea
T^a_{~i}, && ~~ a=1 \dots n+1,\, i=1 \dots n \non
\mbox{rk}(T) &=& n ~.
\eea
 The $n+1$ rows are linearly dependent, yet any $n$ of them are independent. Therefore there exists a single vector $l_a$ which is in the left hand side null subspace (kernel) of $T$, namely \be
 l_a\,  T^a_{~i} = 0
 \label{l_null1}
 \ee
Moreover, the left null vector $l_a$ can be expressed in terms of $T$ as follows \be
l_a =  \eps_{a a_1 \dots a_n}\, \eps^{i_1 \dots i_n}\,  T^{a_1}_{~i_1} \dots T^{a_n}_{~i_n}
\label{def:null1}
\ee
 This expression tells us that component $a$ of $l$ is gotten by erasing row  $a$ in $T$, computing the determinant (or minor) of the resulting square matrix, and finally multiplying by an appropriate sign. 
 
To see why definition (\ref{def:null1})  satisfies (\ref{l_null1}) note that since each component of $l_a$ is given by an $n$ by $n$ determinant then for any column $i$ of $T$ $l_a\,  T^a_{~i}$ represents an expansion of an $n+1$ by $n+1$ determinant gotten by joining with $T$ its column $i$. Since this matrix has a repeated column its determinant vanishes thereby proving the point. 

To motivate  (\ref{def:null1}) consider normalizing one of the component of $l_a$ to unity, say $l_{n+1}=1$. Now the remaining components are determined and can be expressed through Cramer's rule as a ratio of minors. Multiplying this vector by the common denominator (the minor gotten by erasing row $n+1$ of $T$) we arrive at (\ref{def:null1}).

Summarizing this first warm up example, it shows how maximal minors can be used to represent the null subspace of a matrix. 

\presub {\bf Example 2}. As a second warm up example consider an $n+1$ by $n+1$ matrix $T^a_{~i}$ of (sub-maximal) rank $n$, namely  \bea
T^a_{~i}, && ~~ a=1 \dots n+1,\, i=1 \dots n+1 \non
\mbox{rk}(T) &=& n ~.
\eea
Given its rank there should be a single left null vector $l_a$ and a single right null vector $r^i$ ($l_a,\, r^i$ are unique up to rescaling), namely \be
  l_a\,  T^a_{~i} =0 \qquad  T^a_{~i}\, r^i =0 
 \ee

In order to find expressions for $l_a$ and $r^i$ one proceeds as follows. One defines the adjugate of $T$,  $M$, such that each of its components $M^i_{~a}$ is the minor gotten by erasing row $a$ and column $i$ of $T$ followed by multiplication by the sign $(-1)^{i+a}$. Equivalently \be
M^i_{~a} =  \eps^{i i_1 \dots i_n}\,  \eps_{a a_1 \dots a_n}\, T^{a_1}_{~i_1} \dots T^{a_n}_{~i_n} ~~.
\label{def:null2}
\ee

For $T$ of any rank we have \bea
T^a_{~i} \, M^i_{~b} &=& \det(T) \delta^a_{~b} \label{TM} \\
M^i_{~a} \, T^a_{~j} \,  &=& \det(T) \delta^i_{~j} \label{MT}
\eea
When $\det(T) \neq 0$ the adjugate matrix leads to the inverse matrix through $T^{-1}=M / \det(T)$. However, we assumed that $\mbox{rk}(T)=n$ and hence $\det(T)=0$. (\ref{TM}) implies that each column $b$ of $M$ is in the right null subspace. Since all right null vectors are proportional to $r^i$ we must have $M_i^{~a} = r^i\, \tilde{l}_a$ where each component of $\tilde{l}_a$ represents a possible rescaling factor and together they form the components of some row vector $\tilde{l}$. Similarly, from (\ref{MT}) one deduces that $M^i_{~a}$ must be of the form $M^i_{~a} = \tilde{r}^i\, l_a$ for some column vector $\tilde{r}^i$. Together (\ref{TM}, \ref{MT}) imply that \be
M^i_{~a} =c\, r^i\, l_a
\label{Factorization1}
\ee
where $c$ is some non-zero constant.  This tells us that the adjugate matrix, which is a matrix composed out of maximal minors, necessarily factorizes into a right null vector times a left null vector, and this is the purpose of this example.

\subsection{Factorization of maximal minors}

After the warm up we proceed to the general case of a matrix of general size, $m$ by $n$, and a general rank $r$, namely 
\bea
T^a_{~i}, && ~~ a=1 \dots m,\, i=1 \dots n \non
\mbox{rk}(T) &=& r ~.
\label{def:T}
\eea
 
Motivated by the previous examples we define \be
M^I_{~A} :=  \eps^{I i_1 \dots i_r}\,   \eps_{A a_1 \dots a_r}\, T^{a_1}_{~i_1} \dots T^{a_r}_{~i_r} ~~ \\
\label{def:M}
\ee
where $I=(i_{r+1} \dots i_n)$ and $A=(a_{r+1} \dots a_m)$ are multi-indices. The components of $M$ are the maximal minors of $T$. The following theorem describes the special properties of $M^I_{~A}$. It is the main point of the current section and will be useful in the next.

\presub {\bf Maximal minor factorization theorem}. Given any matrix $T^a_{~i}$ as in (\ref{def:T}) its associated maximal minors $M^I_{~A}$ defined in (\ref{def:M}) satisfy \begin{enumerate}

\item[(i)]  $M$ is null, namely  \bea
T^a_{~i} \, M^{i I'}_{~B} &=& 0 \label{M_r_null} \\
M^I_{~a A'} \, T^a_{~j} \,  &=& 0 \label{M_l_null}
\eea
where $I'$ is a multi index with $n-r-1$ indices and $A'$ has $m-r-1$ indices. Since $M^I_{~A}$ is anti-symmetric in both $I$ and $A$ the contracted index can be interchanged with any other.

\item[(ii)] $M$ factorizes as \be
M^I_{~A} =c\, r^I\, l_A
\label{Factorization}
\ee
where $r^I$ and $l_A$ are antisymmetric and unique up to an overall multiplicative scalar.

\end{enumerate}

This theorem generalizes example 1 which demonstrated the minors to be null, and example 2 which demonstrated factorization. For completeness we include a proof.

\presub {\bf Proof}. $M$ is seen to be null by an argument analogous to that in example 1, namely the product $M \cdot T$ represents the expansion of the wedge product of $r+1$ columns of $T$, which must vanish given that $\mbox{rk}(T)=r$ (and similarly for $T \cdot M$).

An alternative proof for the null property would be to note that for minors that are not necessarily maximal multiplication by $T$ gives larger minors through the determinant expansion formula as follows \bea
T^a_{~i} \, M^{i I'}_{~b B'} &=& \delta^a_{[b}\,  M^{I'}_{~B']} \non
M^{i I'}_{~a A'} \, T^a_{~j} \,  &=& \delta^{[i}_{~j}\, M^{I']}_{~A'} ~.
\eea
Now, once we assume $M$ to be maximal the larger minors all vanish and we recover  (\ref{M_r_null},\ref{M_l_null}).

In order to see why $M$ factorizes, recall that $T$ defines a right null subspace of dimension $m-r$. It can be described as the span of a set of independent vectors $r_1^{i_1}, \dots, r_{m-r}^{i_{m-r}}$. However, this description is considerably redundant as it can be replaced by any other spanning set. An alternative way to identify a subspace is through the wedge product \be
r^I \equiv r^{i_1 \dots i_{m-r}} := \bigwedge_{j=1}^{m-r} r_j^{i_j}   ~.
\ee
 The wedge product defines an antisymmetric tensor (or multi-vector) $r^{i_1 \dots i_{m-r}}$ which can be called the null right tensor. It satisfies \be
T^a_{~i}\, r^{i_1 \dots i_{m-r-1} i} = 0
\ee
and it is the only such tensor which satisfies this equation up to an overall multiplicative factor. In this sense it characterizes the subspace in a unique and hence non-redundant way. 

The uniqueness of of the right null tensor together with right null property of $M^I_{~B}$ for any value of the multi-index $B$, as described in (\ref{M_r_null}), implies that $M^I_{~B} = r^I\, \tilde{l}_B$ for some tensor $\tilde{l}_B$. Repeating the argument for the left null space, in analogy with example 2, implies the argued factorization, namely (\ref{Factorization}), where $l_A$ is the left null tensor of $T$, and $c$ is a free constant whose value depends on the chosen normalizations for $r^I$ and $l_A$.

This completes the proof of the maximal minor factorization theorem. We now proceed to a discussion of several related comments.

\presub {\bf Matrix of polynomials}. Below we shall be interested in a  matrix  $T=T(x)$ whose entries are polynomials in some variables denoted collectively by $x$. In this case each component of $M^I_{~A}$ is a polynomial in $x$ and one can be more specific about the choice of the scalar $c$:  it can be chosen to be the greatest common divisor of all the polynomials  $M^I_{~A}(x)$, namely \be
c(x) = \gcd \( M^I_{~A} (x) \) ~.
\label{def:cx}
\ee
 $c(x)$  is unique up to multiplication by a number (a field element).

\presub {\bf Gauss elimination}. The computation of each component of the maximal minor tensor $M^I_{~A}$ requires to evaluate a determinant. It is well known that evaluating determinants through their definition is computationally costly and a more efficient method is provided by the Gauss elimination method where through elementary operations on rows (or columns) the matrix $T^a_{~i}$ can be brought into an upper triangular form. This implies that $L^{-1}\, T = U$ where $U$ is the upper triangular form, and $L^{-1}$ is a lower triangular matrix which records all the row operation carried on $T$. Since the row operations are invertible one also has \be
T= L\, U ~,
\label{LU_decomp}
\ee
 which is known as an ``LU decomposition of $T$''. 

The LU decomposition above is useful since $T$ and $U$ share the same right null subspace, yet the triangular form of $U$ makes it immediate to determine it.

Similarly Gauss elimination can be applied to columns to obtain (a possibly different) LU decomposition. Column operations produce a matrix $L$ of the same size as $T$ together with a square $U$, while row operations produce the opposite sizes, and hence if T is non-square the two decompositions are necessarily different.

Altogether, minors and null subspaces can be computed either directly from the definitions (\ref{def:M},\ref{Factorization}) or by using an LU decomposition (which is essentially Gauss elimination). The choice of method depends on computational convenience. When applying an LU decomposition to a matrix of polynomials the $L,U$ factor generically would become rational (a ratio of polynomials0, yet in such a way that the minors remain polynomial.

\presub {\bf Dualization}. The definition of the tensor of minors $M^I_{~A}$ in (\ref{def:M}) can be thought to involve two steps -- first a wedge product of $T$ with several copies of itself, followed by a dualization on both the $a$ and $i$ indices. Both steps are performed by the $\eps$ tensors  -- first assuring projection onto the antisymmetric sector and then performing dualization. Here we note that the wedge product is the more essential step, while dualization is convenient in the common case when the $r \equiv \mbox{rk} (T)$ is close to either $m$ or $n$.

\section{Reduction of a Feynman Integral}
\label{sec:reduction}

After reviewing the factorization of maximal minors, we proceed to apply it to the Symmetries of Feynman Integrals (SFI) method \cite{SFI}. 

\presub {\bf Set-up}.  SFI considers a Feynman diagram as a function of its most general possible parameters, namely the masses and the kinematical invariants, and a general spacetime dimension $d$. The parameter space is denoted by $X$ and each diagram is associated with set of partial differential equations in $X$. 

Schematically, the SFI equations are of the form \be
c^a\, I + (T^a)^j_{~i}\, x_j\, \del^i\, I = J^a
\label{SFI_schematic}
\ee
where $a=1, \dots , \dim(G)$ labels each equation in the set, $c^a=c^a(d)$ are constants (namely, are independent of $X$),  the matrices $(T^a)^j_{~i}$ define a representation of a group $G$ on $X$, the range of $i,j$ is given by $i,j = 1, \dots, \dim(X)$, $\del^i=\del/\del x_i$ and finally  $J^a$ are terms composed of simpler diagrams. The group $G$ is called the SFI group and it is defined by the topology of the diagram in a natural way.

Here we focus on the differential term $(T^a)^j_{~i}\, x_j\, \del_i\, I$ which is fully defined by the representation of $G$ on $X$. We define the matrix $Tx$  by \be
 \( Tx \) ^a_{~i} := (T^a)^j_{~i}\, x_j
 \label{def:Tx}
 \ee
This matrix will be the object of our study and it will correspond to the general matrix $T^a_{~i}$ in the previous section. The size of the matrix $Tx$ is given by $m=\dim(G)$ by $n=\dim(X)$.

\presub {\bf Factorization}. At any specific point in $x \in X$ we may determine the rank of $Tx$ which equals the dimension of the tangent space to the G-orbit at $x$ and hence to the dimension of this G-orbit
\be
 r(x) := {\rm rk}(Tx) \equiv \dim \( {\rm G-orbit} (x) \) ~.
\ee
The we evaluate the maximal minors \be
M^I_{~A}(x) := \eps_{A a_1 \dots a_r}\, \eps^{I i_1 \dots i_r}\,   Tx^{a_1}_{~i_1} \dots Tx^{a_r}_{~i_r} ~~ \\
\label{def:Mx}
\ee
(gotten by substituting $T \to Tx$ in the general definition  (\ref{def:M}) ).

Factorization of maximal minors (\ref{Factorization}) in the presence of polynomials (\ref{def:cx}) implies \be
M^I_{~A}(x) =S(x)\, Orb^I(x)\, Stb_A(x)
\label{MxFactor}
\ee
where all the terms are defined up to a multiplicative $x$-independent constant. The notation reflects the interpretation of the various terms in the context of $Tx$  as we proceed to explain.

$S(x)$ is the factor common to all minors, denoted by $c(x)$ in  (\ref{Factorization},\ref{def:cx}). At zeroes of $S(x)$ all the minors vanish reflecting a drop in the rank of $Tx$. We refer to this zero locus as the singularity locus, and accordingly the notation $S(x)$ was chosen to stand for singular.

$Orb^I(x)$ is the right null tensor, denoted by $r^I$ in  (\ref{Factorization}). This null subspace is composed of cotangent vectors at $x$ which annihilate all the rows in $Tx$ and hence annihilate (are perpendicular to) the orbit of $G$ through $x$.  The notation $Orb$ refers to this relation with the orbit. By its definition, the orbit tensor includes the differentials of all group invariants, namely $Orb \wedge dInv=0$ where $Inv$ is any invariant of $G$.

$Stb_A(x)$ is the left null tensor, denoted by $l^A$ in  (\ref{Factorization}). Vectors in this subspace represent a combination of equations (rows of $Tx$) whose action  vanishes at $x$ and is hence known as the stabilizer of the group. Correspondingly, the notation $Stb$ stands for stabilizer.  Note that a multiplication of the SFI equation set from the left by a stabilizer vector generates by definition an equation with no derivatives, namely an algebraic rather than differential equation. For some $x$ the inner product of a stabilizing vector $Stb_a$ with $c^a$, the vector of constants in (\ref{SFI_schematic}), is non-zero thereby generating a simple equation for $I$ which yields a reduction of $I$ to a linear combination of simpler integrals \cite{locus}.
 
The preceding interpretation of the terms in (\ref {MxFactor}) is summarized by the following list \\
\begin{tabular}{lcl}
$c(x)$, the common factor for minors  & $\to$ &  $S(x)$ whose zeroes are the singularity locus \\
right null space & 	$\to$ & co-orbit form subspace	\\
left null space & 	$\to$ & Stabilizer subspace. 	
\end{tabular} 

Equation (\ref{MxFactor}) is the main result of this paper, providing a systematic way to compute the singularity locus, the orbit and its invariants and/or the group stabilizer through minors of $Tx$.  Some stabilizers provide a reduction of the diagram under study into a linear combination of simpler ones (as discussed above).  Computationally, an LU decomposition (\ref{LU_decomp}) might be performed to facilitate the evaluation of minors (see also a demonstration in the next section).

Since all terms in (\ref{MxFactor}) are polynomials in the parameters $x$ we obtain a useful relation between the degrees with respect to $x$ \be
 r = {\rm deg}_x\, S + {\rm deg}_x\, Orb + {\rm deg}_x\, Stb
\label{deg_balance}
 \ee
where $r= {\rm deg}_x\, M^I_{~A}(x) ={\rm rk}(Tx) \equiv  \dim \( {\rm G-orbit} (x) \)$. 

\presub {\bf From maximal rank to lower ones}.
Let us proceed to examine in more detail various loci in $X$ which correspond to a given rank of $Tx$, namely G-orbits of various dimensions.

\presub {Maximal rank}. Denote the highest possible rank by $Rk$ namely \be
Rk = \max_x {\rm rk} (Tx) ~.
\label{def:Rk}
\ee 
$Rk$ is the generic rank, namely it is achieved in an open set in $X$.

Clearly $Rk$ is bounded by the size of $Tx$ namely $Rk \le \min \{\dim (G),\dim(X)\}$. If the inequality is exhausted such that $Rk = \dim (G) \le \dim(X)$ then the stabilizer is trivial, namely $Stb =1$ in (\ref{MxFactor}), and $Orb$ is non-trivial telling us about $G$ invariants. Similarly if $Rk=\dim (X) \le \dim(G)$ then $Orb$ is trivial and hence the G-orbit is co-dimension zero, implying maximal effectiveness for SFI, while $Stb$ is non-trivial providing us with algebraic equations (probably an algebraic constraint on the sources).

\presub {Next to maximal rank}. Factorization at maximal rank defines the singularity function $S(x)$.  One proceeds to factorize $S(x) = S_1(x)\, S_2(x) \dots$. Each factor of $S(x)$ defines a component of the singular locus over which the rank is reduced, generically to $Rk-1$.  On any component $c=1,2,\dots$ where $S_c(x)=0$ factorization of maximal minors (\ref{MxFactor}) may be re-applied as follows \be
  M_{c~~\tilde{A}}^{~\tilde{I}}(x) = \hat{S}_c(x)\, Orb_c^{~\tilde{I}}(x)\, Stb_{c~\tilde{A}}(x)  \mod S_c(x)
\label{Factor_at_sing}
 \ee
 where $\tilde{I},\tilde{A}$ are larger multi-indices. This time the minors are smaller, and accordingly the stabilizer and orbit subspaces strictly contain the respective maximal rank subspaces. In addition the factorization holds only at $S_c(x)=0$ or equivalently $\mod S_c(x)$.

\presub {Second next to maximal rank}. It could be interesting to proceed to even lower rank at the locus of intersection of several singularity components and to determine the reduction of the Feynman diagram over there.

\presub {\bf Original motivation and chronology}. The idea for this paper appeared while studying the 2-loop propagator diagram (sometimes called lizard eye or marshmallow) \cite{InProgress}, yet notions closely related to maximal minors had appeared earlier in SFI papers and in a prominent way:  the wedge product in \cite{locus}  and the determinant in \cite{VacSeagull}.

\section{Demonstration}
\label{sec:demon}

\begin{figure}
\centering \noindent
\includegraphics[width=4cm]{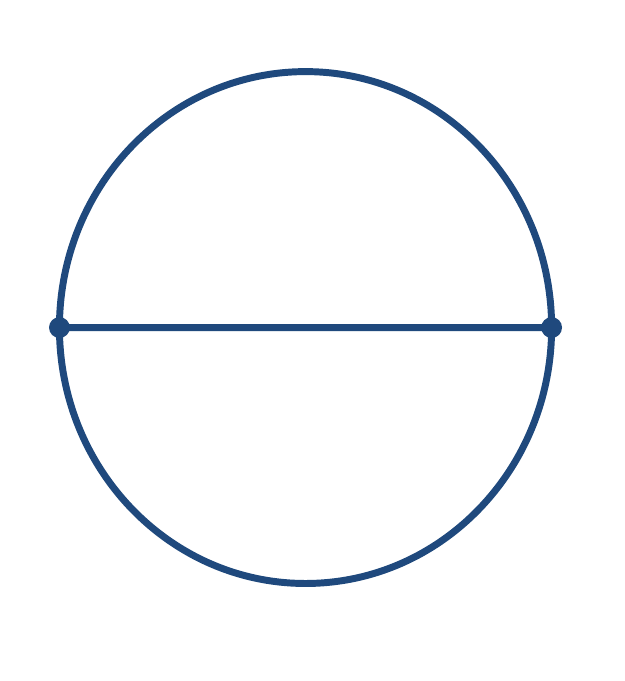}
\caption[]{The diameter diagram.}
\label{fig:diameter}
\end{figure}

In this section we demonstrate the algebraic method of the previous section on the 2-loop vacuum diagram shown in fig. \ref{fig:diameter} which we call the diameter diagram. This provides a simple and non-trivial demonstration.

The parameter space consists of the three possible masses-squared $X=\(x_1,\, x_2,\, x_3 \)$ where $x_i:=(m_i)^2, ~i=1,2,3$. The SFI equation set was found in \cite{SFI}  eq. (6.9) from which we can read \be
Tx = \[ \begin{array}{ccc}
 -x_1		& -x_2		& -x_3 	\\
 x_2-x_3	& x_2		& -x_3 	\\
 -x_1 	& x_3-x_1		& x_3	\\
 x_1 		& -x_2		& x_1- x_2 \\
 \end{array} \] 
 \label{def:Tx_diam}
\ee
There are 4 equations and 3 parameters so $Tx$ is $4$ by $3$.

\presub {\bf Maximal rank}. At a generic point the rank of $Tx$ is 3 and so the G-orbit is 3d. Computing the 3-minors according to (\ref{def:Mx}) we find \be
M_a = \lam  \[  \begin{array}{cccc} 0 & x_1 & x_2 & x_3 \end{array} \]
\label{diam_factor}
\ee
where \be
 \lam := x_1^2 + x_2^2 + x_3^2 - 2 x_1 x_2 - 2 x_1 x_3 - 2 x_2 x_3
\ee
is the Heron formula / K\"all\'en invariant. 

Comparing the expression for the minors with the general factorization (\ref{MxFactor}) we identify \bea
 S &=& \lam \label{s_diam} \\
 Stb &=& \[  \begin{array}{cccc} 0 & x_1 & x_2 & x_3 \end{array} \]
 \label{stb_diam}
 \eea
so the singularities are the zeroes of $\lam$, thereby reproducing \cite{SFI} eq. (6.16) and \cite{locus} eq. (4.9), while the expression for the stabilizer reproduces \cite{locus} eq. (4.12). Multiplying the equation set by this stabilizer produces a relation between sources which is valid everywhere in $X$. The degree balance (\ref{deg_balance}) which corresponds to (\ref{diam_factor})  is \be
3 = 2 +1 ~.
\ee

\presub {\bf LU decomposition}. Alternatively, we may compute the minors and the associated factors through the LU decomposition. To obtain the stabilizer group (null left subspace) we may operate on the right, that is on columns. As a starting point it is convenient to reorder the rows of $Tx$ (\ref{def:Tx_diam}) as \be
\widetilde{Tx}  = \[ \begin{array}{ccc}
  x_1 	& -x_2		& x_1- x_2 \\
   -x_1 	& x_3-x_1		& x_3	\\
 x_2-x_3	& x_2		& -x_3 	\\
  -x_1		& -x_2		& -x_3 	\\
 \end{array} \] 
 \label{Tx_diam_reorder}
\ee
After operating on columns one gets a lower triangular form \be
 \[ \begin{array}{ccc}
  x_1 	& 0		& 0  \\
   -x_1 	& 1		& 0	\\
 x_2-x_3	& -x_2/x_1		& 0	\\
  -x_1		& x_2/x_3		& -\lam 	\\
 \end{array} \] 
\ee
Considering the third row we notice that all 3-minors are proportional to $\lam$ consistent with the singularity factor (\ref{s_diam}). Moreover, from this form a left null vector can be read $\[ \begin{array}{cccc}  x_3/x_1 &  x_2/x_1 & 1  & 0 \end{array} \]$. After multiplying by $x_1$ and reordering to account for the ordering of $\widetilde{Tx}$ we reproduce the stabilizer (\ref{stb_diam}).

For completeness we present also the triangulation of  $\widetilde{Tx}$ through operations on rows. In this case we reach the upper triangular form \be
 U = \[ \begin{array}{ccc}
  x_1 	& -x_2	& x_1-x_2  \\
   0	 	& -2 s^3	& 2 s^2	\\
  0		& 0		& 0	\\
  0		& 0		& -\frac{2}{s^3}\, \lam 	\\ 
 \end{array} \]  ~.
\ee
where we have introduced the notation \be
 s^i := -\frac{\del}{4\, \del x_i}\, \lam 
 \ee 
Only one minor is non-zero -- the one gotten by erasing row 3, and it is indeed proportional to $\lam$ (and the denominator cancels). To find the stabilizer we must record the row operations performed through the lower triangular matrix \be
 L^{-1} = \[ \begin{array}{ccc}
  x_1 	& -x_2	& x_1-x_2  \\
   0	 	& -2 s^3	& 2 s^2	\\
  0		& 0		& 0	\\
  0		& 0		& -\frac{2}{s^3}\, \lam 	\\ 
 \end{array} \] 
\ee
which satisfies $L^{-1}\, \widetilde{Tx} = U$. Now $\tilde{s}=\[ \begin{array}{cccc}  0 &  0 & 1  & 0 \end{array} \]$ is a left null vector for $U$. Multiplication by $L^{-1}$ on its left reproduces the stabilizer (\ref{stb_diam}), after some rescaling and reordering just as before.

\presub {\bf Next to maximal rank}. At the singular/algebraic locus $\lam=0$ the rank of $Tx$ and hence the dimension of the G-orbit reduces to 2 and one computes the maximal minors tensor \be
N^i_{~ab} := \( M_\lam\)^i_{~ab}
\label{def:N}
\ee by omitting column $i$ and rows $a,b$. 

According to the general procedure (\ref{Factor_at_sing})  $N^i_{~ab}$ factorizes. We start by noticing that the co-orbit 1-form $Orb_\lam \equiv Orb$ can be anticipated. By definition at $\lam=const$ the 1-form $d\lam \equiv -4\, s^i \, dx_i$ annihilates all vectors tangent to the locus, and hence we recognize $Orb$ to be
\be
Orb^i  = \[ \begin{array}{c}
s^1 \\
s^2 \\
s^3 \\
\end{array} \]
\label{Orb_lam}
\ee

Next, $N^{i}_{~ab}$ should be divided by $Orb$ to yield $Stb_\lam$. Since this factorization holds for $\lam=0$, but not for all $x$, some more algebra is required.  Factoring, say, $M^3_{~ab}$ by $s^3 \mod \lam$ can be done by eliminating one the coordinates, for instance $x_3$, in terms of $x_1,\, x_2$ but this introduces square roots and makes the algebra awkward. Instead one can work $\mod \lam$ and write $M^3_{~ab}(x)=s^3\, Stb_{\lam~ab}(x) + \lam\, k_{ab}(x)$ where $k_{ab}$ is some matrix. Now it is easier to impose $s^3=0$ by substituting $x_3 \to x_1+x_2$. This allows to determine $k_{ab}$ which happens to be a matrix of constants, and now $ Stb_{\lam}$ can be determined to be
\be
 Stb_{\lam~ab} = \[ \begin{array}{cccc}
  0 		& x_1	& x_2	& x_3 	 \\
  -x_1	&0	 	& s^3	& -s^2	\\
  -x_2	& -s^3	& 0		& s^1 	\\
  -x_3	& s^2	& -s^1 	& 0    	\\ 
 \end{array} \]  ~.
\label{Stb_lam}
\ee

Since at $\lam=0$ ${\rm rk} \( Tx \) =2$ the stabilizer subgroup is 2d and hence the tensor $Stb_\lam$ tensor has rank 2, namely it is a bi-vector. Its first row confirms that it includes $Stb$ (\ref{stb_diam}), the general stabilizer which is valid for all $x$ and leads to the algebraic constraint for the sources. 

Any of the remaining three rows can be used to obtain the algebraic solution, namely the reduction of the diameter to simpler diagrams. Their sum reproduces eq. (4.10) of \cite{locus}. However, we notice that if we pick one of them, say the 1st, it becomes apparent that $\mod \lam$ the expression for the algebraic solution  simplifies to \be 
\left. I \right|_\lam = \frac{1}{d-3} \[ s^3\, j'(x_1)\, j'(x_2) + cyc. \] 
\label{I_lam}
\ee
where the notation is the same as in \cite{locus}. The simplification occurs due to a non-manifest cancellation of the denominator $\mod \lam$ in eq. (4.11). 

Summarizing the factorization at $\lam=0$ we have \be
N^i_{~ab}  = Orb^i\,  Stb_{\lam~ab}  \mod \lam
\label{N_factor}
\ee
where $Orb$ is given in (\ref{Orb_lam}) and $\left. Stb_{\lam~ab} \right|_\lam$ in (\ref{Stb_lam}). There is no non-trivial scalar common factor so $ \hat{S}_\lam(x) =1$. The degree balance reads \be
2 = 1 + 1 ~.
\ee

\section{Summary and Discussion}
\label{sec:disc}

In this paper we analyzed certain algebraic aspects of the SFI equation set, and showed that factorization of maximal minors is useful to determine  the singular locus, the G-orbit together with the G-invariants and the stabilizer (\ref{MxFactor}). On some orbits the latter provides a reduction of the diagram under study into a sum of simpler diagrams. Factorization can be performed over any G-orbit, including those whose dimension is lower than the generic value.

The method was illustrated through the diameter diagram. 

We end with two comment. First, at the algebraic  locus we have an exact solution. It would be interesting to develop a perturbation theory in its vicinity. 

Secondly, the procedure described in this paper, depends only on the representation of $G$ on $X$ and it involves minors which are antisymmetric and hence suggest fermionic variables. In these respects it is similar to Group Cohomology, where the ghost and anti-ghosts are fermionic. It would be interesting to find out whether this similarity is more substantial.

\subsection*{Acknowledgments}

It is a pleasure to thank Subhajit Mazumdar, Lior Oppenheim, Amit Schiller and Ruth Shir and for collaboration on related projects and for comments on a presentation.

This research was supported by the ``Quantum Universe'' I-CORE program of the Israeli Planning and Budgeting Committee.

\bibliographystyle{unsrt}

\begin{thebibliography}{99}

\bibitem{SmirnovBooks} 
  V.~A.~Smirnov,
  ``Feynman integral calculus,''
  Berlin, Germany: Springer (2006). 
  V.~A.~Smirnov,
  ``Analytic tools for Feynman integrals,''
  Springer Tracts Mod.\ Phys.\  {\bf 250}, 1 (2012).

\bibitem{ChetyrkinTkachov1981} 
  K.~G.~Chetyrkin and F.~V.~Tkachov,
  ``Integration by Parts: The Algorithm to Calculate beta Functions in 4 Loops,''
  Nucl.\ Phys.\ B {\bf 192}, 159 (1981).
%

\bibitem{DE}
  A.~V.~Kotikov,
  ``Differential equations method: New technique for massive Feynman diagrams calculation,''
  Phys.\ Lett.\ B {\bf 254}, 158 (1991). \\
  E.~Remiddi,
  ``Differential equations for Feynman graph amplitudes,''
  Nuovo Cim.\ A {\bf 110}, 1435 (1997)
  [hep-th/9711188]. \\
  T.~Gehrmann and E.~Remiddi,
  ``Differential equations for two loop four point functions,''
  Nucl.\ Phys.\ B {\bf 580}, 485 (2000)
  [hep-ph/9912329].


 \bibitem{SFI}  
  B.~Kol,
  ``Symmetries of Feynman integrals and the Integration By Parts method,''
  arXiv:1507.01359 [hep-th].

 \bibitem{locus} 
  B.~Kol,
  ``The algebraic locus of Feynman integrals,''
  arXiv:1604.07827 [hep-th].

\bibitem{bubble} 
  B.~Kol,
  ``Bubble diagram through the Symmetries of Feynman Integrals method,''
  arXiv:1606.09257 [hep-th].

\bibitem{VacSeagull} 
\bibitem{Burda:2017tcu} 
  P.~Burda, B.~Kol and R.~Shir,
  ``Vacuum seagull: Evaluating a three-loop Feynman diagram with three mass scales,''
  Phys.\ Rev.\ D {\bf 96}, no. 12, 125013 (2017)
  doi:10.1103/PhysRevD.96.125013
  [arXiv:1704.02187 [hep-th]].
  
\bibitem{InProgress}
 B.~Kol and S.~Mazumdar,
 ``Marshmallow diagram through the Symmetries of Feynman Integrals method,''
 in progress.  
  
\end{thebibliography}

\end{document}